\documentclass{article}

\usepackage{arxiv}

\usepackage[utf8]{inputenc} 
\usepackage[T1]{fontenc}    
\usepackage[hidelinks]{hyperref}       
\usepackage{url}            
\usepackage{booktabs}       
\usepackage{amsfonts}       
\usepackage{nicefrac}       
\usepackage{microtype}      
\usepackage{amsmath}
\usepackage{cleveref}       
\usepackage{lipsum}         
\usepackage{caption} 
\usepackage{subcaption}     
\usepackage{tabularx}       
\usepackage{booktabs}       
\usepackage{float}          
\usepackage{calc}           
\usepackage{blindtext}
\usepackage{multirow}
\usepackage{csquotes}       
\usepackage[inkscapelatex=false]{svg} 
\usepackage{setspace}
\usepackage{changepage} 
\usepackage[breakable]{tcolorbox}
\usepackage{enumitem}
 \usepackage{multirow}
\usepackage{colortbl}
\usepackage{abstract}

\usepackage[paper=portrait,pagesize]{typearea}  

\usepackage{graphicx}
\usepackage{natbib}
\usepackage{doi}

\title{Applying Pre-Trained Deep-Learning Model
on Wrist Angel Data - An Analysis Plan}

\newif\ifuniqueAffiliation
\uniqueAffiliationtrue

\ifuniqueAffiliation 
\author{ 
Harald Vilhelm Skat-Rørdam 
\thanks{These authors contributed equally} 
\thanks{Applied Mathematics and Computer Science, Technical University of Denmark, Kongens Lyngby, Denmark} \\
	\texttt{s175393@dtu.dk} \\
	\And
 Mia Hang Knudsen
\footnotemark[1] \footnotemark[2] \\
	\texttt{s183998@dtu.dk} \\
 	\And
 Simon Nørby Knudsen
 \footnotemark[1] \footnotemark[2] \\	\texttt{s174479@dtu.dk} \\
\AND
 Nicole Nadine Lønfeldt \thanks{Child and Adolescent Mental Health Center, Copenhagen University Hospital, Mental Health Services Copenhagen, Hellerup, Denmark}\\	\texttt{nicole.nadine.loenfeldt@regionh.dk} 
 \And
 Sneha Das \footnotemark[2] \\	\texttt{sned@dtu.dk} 
 \\
   	\AND
 Line Katrine Harder Clemmensen \footnotemark[2] \\	\texttt{lkhc@dtu.dk} 
 \\
}
\else
\usepackage{authblk}

\newbox{\orcid}\sbox{\orcid}{\includegraphics[scale=0.06]{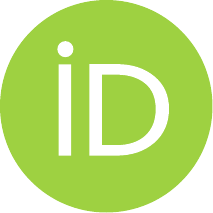}} 
\author[1]{%
	\href{https://orcid.org/0000-0000-0000-0000}{\usebox{\orcid}\hspace{1mm}David S.~Hippocampus\thanks{\texttt{hippo@cs.cranberry-lemon.edu}}}%
}
\author[1,2]{%
	\href{https://orcid.org/0000-0000-0000-0000}{\usebox{\orcid}\hspace{1mm}Elias D.~Striatum\thanks{\texttt{stariate@ee.mount-sheikh.edu}}}%
}
\affil[1]{Department of Computer Science, Cranberry-Lemon University, Pittsburgh, PA 15213}
\affil[2]{Department of Electrical Engineering, Mount-Sheikh University, Santa Narimana, Levand}
\fi

\renewcommand{\shorttitle}{Statistical Analysis Plan}

\hypersetup{
pdftitle={Applying Pre-Trained Deep-Learning Model
on Wrist Angel Data - An Analysis Plan},
pdfsubject={q-bio.NC, q-bio.QM},
pdfauthor={Harald Vilhelm Skat-Rørdam, Mia Hang Knudsen, Simon Nørby Knudsen},
pdfkeywords={Deep Learning, OCD, Predict},
}

\begin{document}
\maketitle

\begin{abstract}
     We aim to investigate if we can improve predictions of stress caused by OCD symptoms using pre-trained models, and present our statistical analysis plan in this paper. With the methods presented in this plan, we aim to avoid bias from data knowledge and thereby strengthen our hypotheses and findings.
    
    The Wrist Angel study, which this statistical analysis plan concerns, contains data from nine participants, between 8 and 17 years old, diagnosed with obsessive-compulsive disorder (OCD). The data was obtained by an Empatica E4 wristband, which the participants wore during waking hours for 8 weeks. The purpose of the study is to assess the feasibility of predicting the in-the-wild OCD events captured during this period. 
    
    In our analysis, we aim to investigate if we can improve predictions of stress caused by OCD symptoms, and to do this we have created a pre-trained model, trained on four open-source data for stress prediction. We intend to apply this pre-trained model to the Wrist Angel data by fine-tuning, thereby utilizing transfer learning. 
    The pre-trained model is a convolutional neural network that uses blood volume pulse, heart rate, electrodermal activity, and skin temperature as time series windows to predict OCD events. 
    Furthermore, using accelerometer data, another model filters physical activity to further improve performance, given that physical activity is physiologically similar to stress. 
    
    By evaluating various ways of applying our model (fine-tuned, non-fine-tuned, pre-trained, non-pre-trained, and with or without activity classification), we contextualize the problem such that it can be assessed if transfer learning is a viable strategy in this domain.  
\end{abstract}


\newpage
\section{Scope of analysis}
\label{sec:scope}
This analysis aims to evaluate whether transfer learning is a viable option for enhancing OCD-event predictability, which could be generalized to other areas of interest. This is especially relevant for new prediction tasks, where data scarcity is problematic.

This is done by assessing if a pre-trained deep-learning model can improve the performance of predicting obsessive-compulsive disorder (OCD) events compared to traditional machine learning techniques that were previously applied \cite{Line_resultater}, using the same data set, from here on referred to as Wrist Angel data. \\
A proposed daily plan for data analysis on Wrist Angel data can be seen in \cref{app:daily_plan}.

The primary questions this analysis aims to answer are:
\begin{enumerate}
    \item To what extent does deep learning improve performance compared to traditional machine learning techniques?
    \item How might the performance of a pre-trained model (with and without fine-tuning) compare to that of a non-pre-trained model?
    \item What is the influence of using a simple activity model on stress prediction?
    \item 
    How does predicting OCD events at various time intervals, such as 0, 1, 2, 3, 4, and 5 minutes before the event, affect performance? This interval between prediction and event will be referred to as prediction lead time. 
    \item How does the length of the signal sequence influence model performance?
\end{enumerate}

\begin{figure}[H]
    \centering
    \includegraphics[width=1\textwidth]{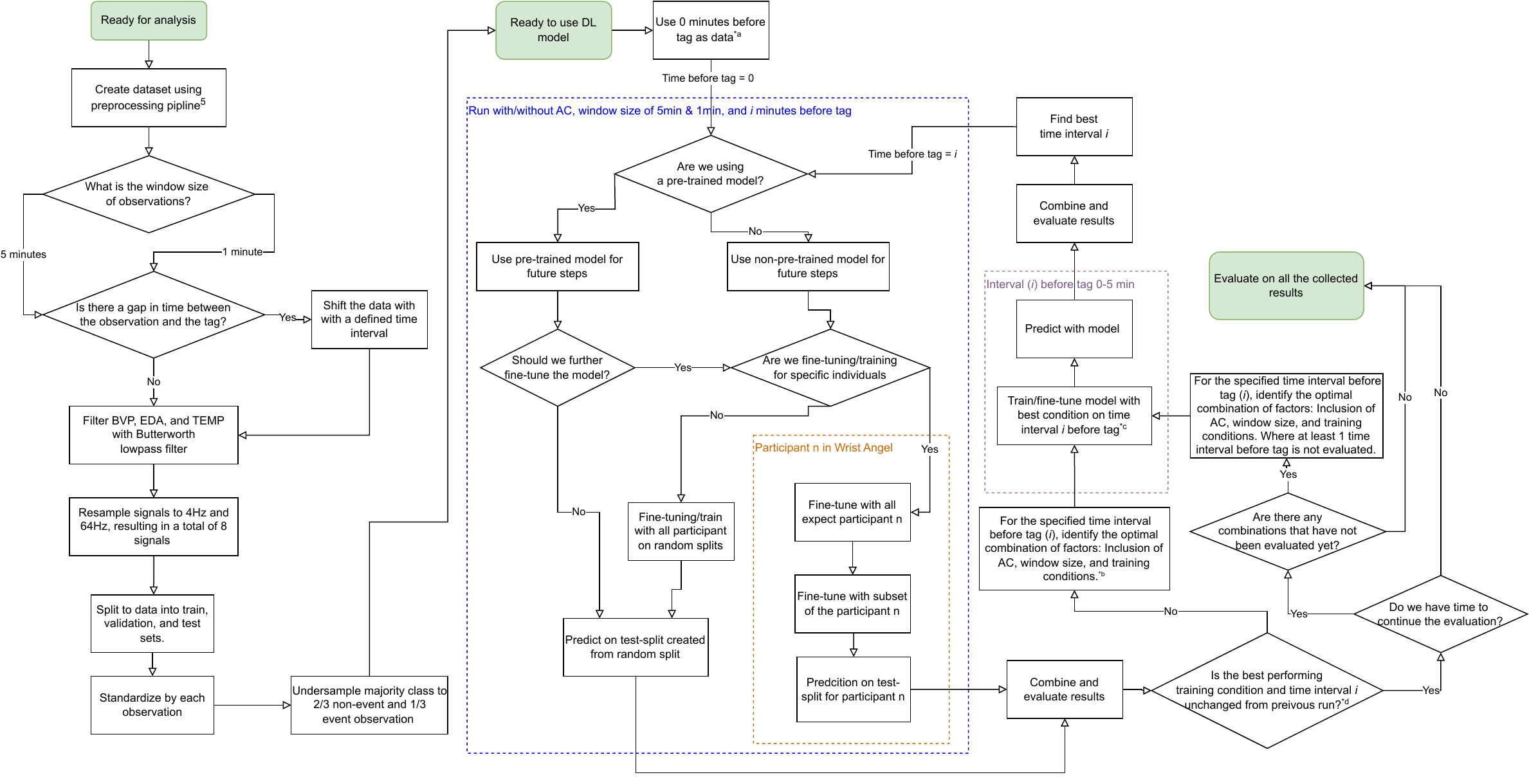}
    \caption{A flow chart describing the steps in signal processing and data analysis. The performance of models will be logged in \cref{tab:model_runs}. \textit{(All figures are PDF).}\\
    Abbreviations. BVP: Blood volume pulse, EDA: Electrodermal activity, HR: Heart rate, TEMP: Skin temperature, DL: Deep learning, AC: Activity classification.\\
     $^*$ For a visual example of the optimization steps see \cref{tab:FlowTable} in Appendix. The letters correspond to the sub-tables.\\
    $^5$ The preprocessing pipeline is the same as used in \cite{Line_resultater}.}
    \label{fig:flowChart}
\end{figure}

\section{Proposed Data Preprocessing}
To get an understanding of the data, we will perform an exploratory analysis, computing metrics such as means and standard deviations for continuous variables and frequencies for categorical variables. Furthermore, we will plot histograms for physiological and accelerometer data. 
The accelerometer data will be used to evaluate and tune the threshold for a physical activity classification. 

The preprocessing pipeline will be the same one used in the previous study and is explained in \cite{Line_resultater} with the possibility of minor changes as explained below. 

The data used in this study comes from the E4 wristband, 4 signals will be used in our model: Blood volume pulse (BVP), heart rate (HR), electrodermal activity (EDA), and skin temperature (TEMP). The tags and the timestamp related to the tag are used to pinpoint the stress episodes and to create the dataset. 
Additionally, accelerometer data is used for our physical activity classification. 

We will extract signals using the window strategy and test different window sizes of 60 seconds and 5 minutes. There will be a 5-minute buffer after the tag where no data will be extracted. There will be no overlapping signals. 


Experiments will be done with different prediction lead times to see how early we can predict an OCD event. Specifically, prediction lead times of 0, 1, 2, 3, 4, and 5 minutes. These are practically gaps between the end of the signal and the OCD event tag. 
It is done by extracting signals ending at the same time as the tag and increasing the interval up to 5 minutes with a 1-minute interval, resulting in a total of the above-mentioned 6 different ways of extracting event signals before the tag. 

A sixth-order Butterworth low-pass filter with a cut-off frequency of 1Hz will be applied to EDA and TEMP. For BVP a second-order Butterworth filter with a high cut-off frequency of 12 Hz and a low cut-off frequency of 2 Hz will be used, while HR will remain unprocessed.

We plan to re-sample the data to obtain the same number of samples for all signals. Both down-sampled (4Hz) and up-sampled (64Hz) data will be used.
BVP and EDA will be re-sampled using python \texttt{scipy.signal} which uses a Fourier method. HR and TEMP are re-sampled using linear interpolation. 

The final step shall be splitting the data and standardizing the signals. This will be done by first splitting the data into train, validation, and test sets. The selected standardization method is observation-wise, meaning, we will standardize the data one observation at a time, each signal individually in all three splits. 

To deal with the unbalanced data, under-sampling is used, by randomly selecting non-event windows from all non-event data. The final distribution of our data will be 1/3 event data and 2/3 non-event data.

\section{Data sets used to pre-train our model}
We have pre-trained our deep learning model on four different open-source data sets.
The datasets used are:
\begin{enumerate}
    \item The DTU dataset 
    \\
    It is collected through an experiment conducted at DTU. A total of 28 participants spread over three cohorts participated in the experiment. All participants wore an Empatica E4 wristband. The experiment consisted of three 5-minute phases. First, a pre-task resting phase, then, the task-solving phase, and finally, a post-task resting phase. The timed puzzle task was categorized as a stress event.
    
    \item WESAD \cite{wesad} \\
    WESAD is a dataset combining physiological data with stress exposure. This dataset contains data from 15 participants, each wearing an Empatica E4.
    The participants went through a stressful scenario, where they were exposed to the Trier Social Stress Test (TSST)\cite{tsst}. The stress test consists of a 5-minute speech and a mental arithmetic task. The non-stressful category comes from scenarios of resting at a table and watching funny video clips. 
    
    \item AffectiveROAD dataset \cite{Road_dataset} \\
    This is from an experiment with 14 drives, from 10 different participants wearing two Empatica E4s. All participants drive the same route for three different types of roads, being exposed to stressful and non-stressful conditions. The "stress" metric is created based on a subjective measure from a co-driver, and the driver itself.  
    
    \item ADARP dataset \cite{adarp_dataset}\\
    A dataset was created by Washington State University, to examine if it is possible to detect stress in subjects diagnosed with alcohol use disorder (AUD). The data was collected from 11 participants over an average of 7 days, where they were asked to wear an E4 wristband during their daily life and tag times when they felt stressed.  
\end{enumerate}

\section{Deep Learning Modeling}
\label{sec:DLmodel}
The event classification model we wish to apply to the Data from the Wrist Angel Feasibility Study consists of a 1D convolutional auto-encoder, from which we use the encoder and attach it to a convolutional classification network. Details about the model can be seen in \cref{app:model}. This model will be used both as a pre-trained model and as a non-pre-trained model.

Furthermore, we intend to include a threshold for standard deviation or Fourier transform frequency to classify physical activity based on accelerometer data, before using the event classification model. To determine which of the two classification methods, standard deviation or Fourier transform frequency, will be used, the in-clinic session with dance and relax data is used to create a baseline for when the children are in physical activity. The method that can most effectively distinguish the activity will then be used as our activity classification. 

When using the physical activity classification, all observations classified as physical activity will be removed from the data, before performing any fine-tuning/predicting. 
If 25\% (based on statistics from \cite{mota_patterns_2003}) or more of the data is removed due to the physical activity classification, we will re-sample the data to fine-tune and predict data with less physical activity. 
When the physical activity classification is not used to filter the data, no observations are removed, but will still be classified, such that we can compare the physical activity classification to the final OCD event classification at a later time.

\begin{figure}[H]
\centering
\begin{subfigure}{1\textwidth}
  \centering
  \includegraphics[scale=0.35]{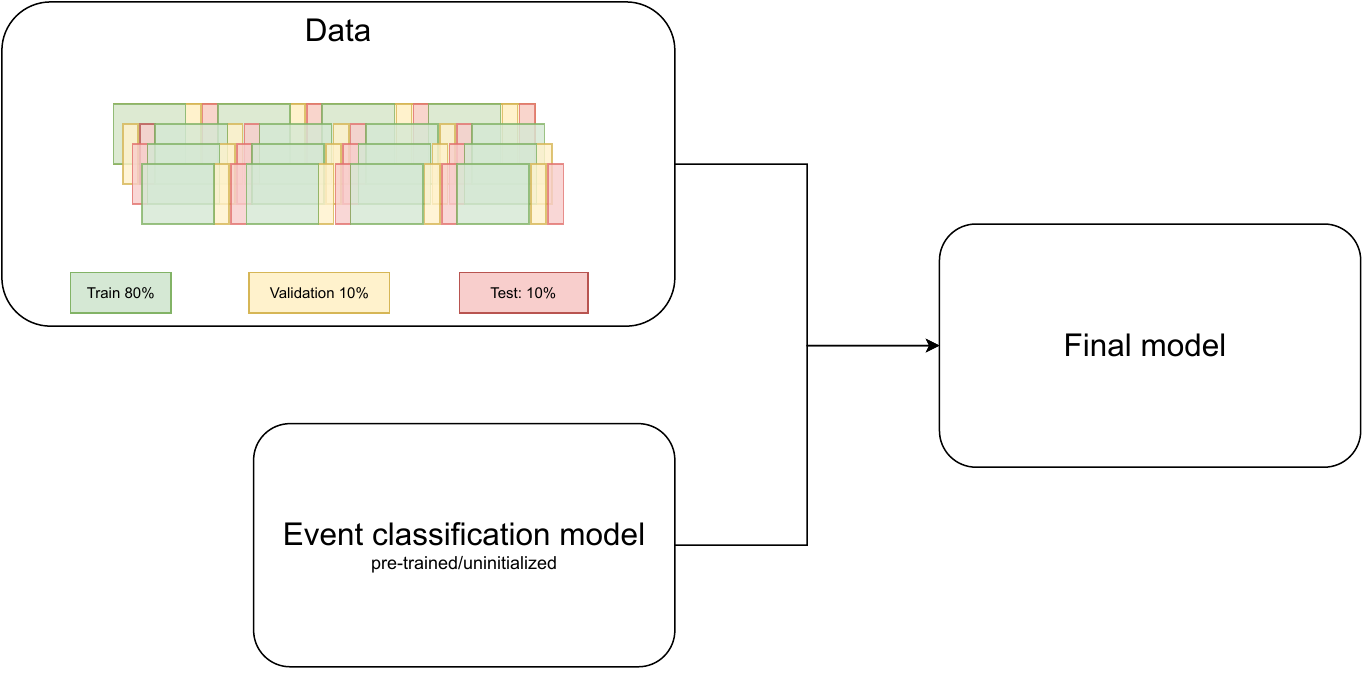}
    \caption{Flow for fine-tuning method: Random.}
    \label{fig:flow2Random}
\end{subfigure}
\begin{subfigure}{1\textwidth}
  \centering
\includegraphics[scale=0.35]{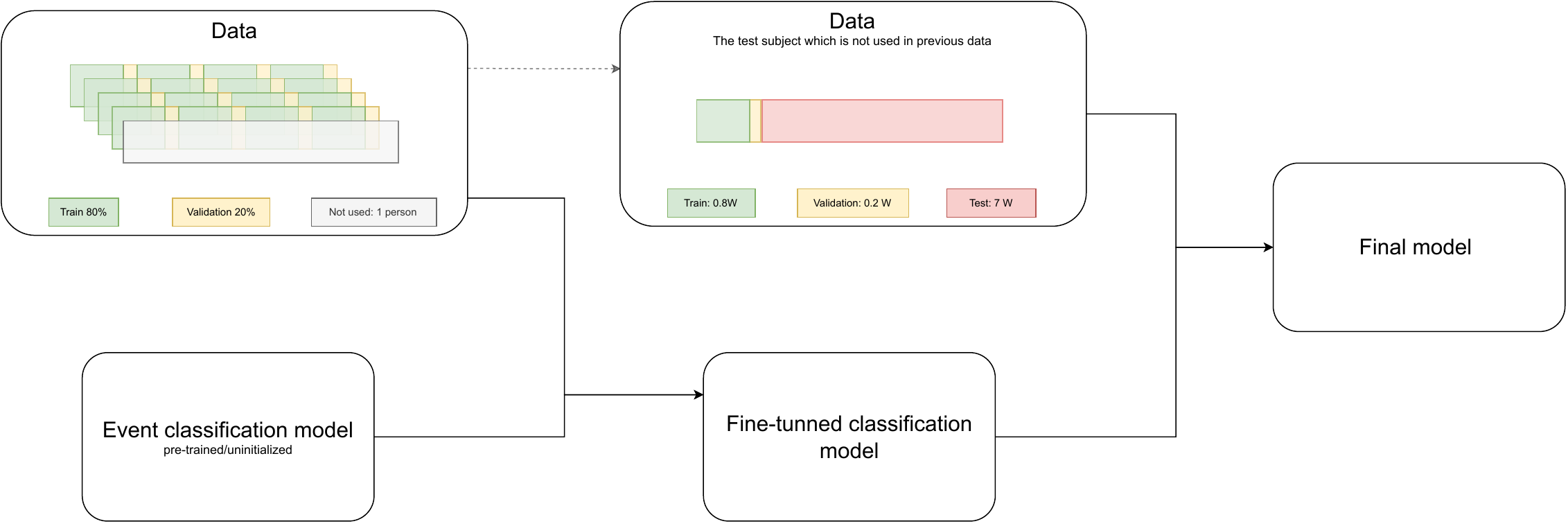}
    \caption{Flow for fine-tuning method: Random + Personalized.}
    \label{fig:flow2}
\end{subfigure}
\caption{Visualization of the two fine-tuning/training methods. }
\label{fig:FlowFineTrain}
\end{figure}

We intend to apply our event classification model in five different ways:
\begin{enumerate}
    \item Apply our pre-trained model directly to the Wrist Angel data to predict OCD events and assess the performance of the model, without the model knowing anything about the specific dataset.
    \item Fine-tune our pre-trained model to a subset of the Wrist Angel data, and predict OCD events of the remaining data. Assess performance, to see if fine-tuning helps. We will fine-tune in two different ways:
    \begin{enumerate}
        \item Random (\cref{fig:flow2Random}): Randomly splitting the Wrist Angel data in 80\% train, 10\% validation and 10\% test.
        \item Random + Personalized (\cref{fig:flow2}): Two fine-tunes. First, we extract one person as the test subject. The remaining 8 persons will be randomly split into 80\% train and 20\% validation data for the first fine-tuning. Thereafter, a second fine-tuning will be applied to the test subject, where the first week will be split into 80\% training and 20\% validation. The remaining 7 weeks will be used for testing. This will then be repeated such that each person has been extracted, giving us a total of 9 tests. 
    \end{enumerate}
    \item Not using our pre-trained model, but training with the Wrist Angel data on our uninitialized model (not pre-trained). This will be used to assess if a pre-trained model improves performance. Training/fine-tuning in the same two ways as explained above with the pre-trained model.
\end{enumerate}

We will apply our event classification model in the stated five different ways, both with and without doing a physical activity classification beforehand. This classification aims at removing time-series windows where the participant has performed physical activity, to not risk predicting it as an OCD event, since physical activity can have similar effects on physiological signals as stress. 

\section{Evaluation methods}
To evaluate our models we will use the metrics accuracy and F1-score.
\begin{equation}
\label{eq:acc}
Accuracy = \frac{TP+TN}{TP+FP+TN+FN} 
\end{equation}
\begin{equation}
\label{eq_f1}
\text{F1-score} = 2 \cdot \frac{Precision \cdot Recall}{Precision + Recall} = \frac{2 \cdot TP}{2 \cdot TP + FP + FN}
\end{equation}

Where $TP$ is true positives, $TN$ is true negatives, $FP$ is false positives, and $FN$ is false negatives. The $F1$-$score$ is the harmonic mean of the two metrics $Precision$ and $Recall$. 

Furthermore, we will plot Receiver Operating Characteristic (ROC) curves to investigate different thresholds for the classifications, and evaluate these using the Area Under the Curve (AUC). 

We will evaluate our model for each of the 5 different ways as explained in \cref{sec:DLmodel}. Additionally, we will evaluate each of these with and without the physical activity classification. Furthermore, we will do everything for 60-second intervals and 5-minute intervals. Finally, we will do everything on the 6 different extracted signals, namely the different periods between the signal and OCD event tag.
In total, this gives us 120 runs, as can be seen in \cref{tab:model_runs}. 
We might not have time to evaluate our model in all these 120 ways. Hence, we intend to optimize the order by first evaluating everything for 0 min prediction lead time, then continue with the additional prediction lead times for the best result, etc. This strategy is visualized in \cref{fig:flowChart}
\begin{equation}
    \label{eq:N_runs}
    \text{Number of runs} = 5 \cdot 2 \cdot 2 \cdot 6 = 120
\end{equation}

\begin{table}[H]
\centering
\resizebox{0.78\textwidth}{!}{%
\begin{tabular}{lclllll}
\multicolumn{1}{c}{} & \multicolumn{6}{c}{Prediction lead time} \\
\multicolumn{1}{c}{\multirow{-2}{*}{\textbf{5min data Physical activity classification}}} & 0min & 1min & 2min & 3min & 4min & 5min \\
{\color[HTML]{000000} Pre-trained - Predict directly (non fine-tuned)} &  &  &  &  &  &  \\
{\color[HTML]{000000} Pre-trained - Random fine-tuned} &  &  &  &  &  &  \\
{\color[HTML]{000000} Pre-trained - Random + personalized fine-tuned} &  &  &  &  &  &  \\
{\color[HTML]{000000} Uninitialized - Random train} &  &  &  &  &  &  \\
{\color[HTML]{000000} Uninitialized - Random + personalized train} &  &  &  &  &  &  \\ \hline
\multicolumn{1}{c}{} & \multicolumn{6}{c}{Prediction lead time} \\
\multicolumn{1}{c}{\multirow{-2}{*}{\textbf{5min data No physical activity classification}}} & 0min & 1min & 2min & 3min & 4min & 5min \\
{\color[HTML]{000000} Pre-trained - Predict directly (non fine-tuned)} &  &  &  &  &  &  \\
{\color[HTML]{000000} Pre-trained - Random fine-tuned} &  &  &  &  &  &  \\
{\color[HTML]{000000} Pre-trained - Random + personalized fine-tuned} &  &  &  &  &  &  \\
{\color[HTML]{000000} Uninitialized - Random train} &  &  &  &  &  &  \\
{\color[HTML]{000000} Uninitialized - Random + personalized train} &  &  &  &  &  &  \\ \hline
\multicolumn{1}{c}{} & \multicolumn{6}{c}{Prediction lead time} \\
\multicolumn{1}{c}{\multirow{-2}{*}{\textbf{1min data Physical activity classification}}} & 0min & 1min & 2min & 3min & 4min & 5min \\
{\color[HTML]{000000} Pre-trained - Predict directly (non fine-tuned)} &  &  &  &  &  &  \\
{\color[HTML]{000000} Pre-trained - Random fine-tuned} &  &  &  &  &  &  \\
{\color[HTML]{000000} Pre-trained - Random + personalized fine-tuned} &  &  &  &  &  &  \\
{\color[HTML]{000000} Uninitialized - Random train} &  &  &  &  &  &  \\
{\color[HTML]{000000} Uninitialized - Random + personalized train} &  &  &  &  &  &  \\ \hline
\multicolumn{1}{c}{} & \multicolumn{6}{c}{Prediction lead time} \\
\multicolumn{1}{c}{\multirow{-2}{*}{\textbf{1min data No physical activity classification}}} & 0min & 1min & 2min & 3min & 4min & 5min \\
{\color[HTML]{000000} Pre-trained - Predict directly (non fine-tuned)} &  &  &  &  &  &  \\
{\color[HTML]{000000} Pre-trained - Random fine-tuned} &  &  &  &  &  &  \\
{\color[HTML]{000000} Pre-trained - Random + personalized fine-tuned} &  &  &  &  &  &  \\
{\color[HTML]{000000} Uninitialized - Random train} &  &  &  &  &  &  \\
{\color[HTML]{000000} Uninitialized - Random + personalized train} &  &  &  &  &  & 
\end{tabular}%
}
\caption{The maximum of 120 ways we intend to apply our model to the Wrist Angel data. Depending on time restrictions we might not complete everything. We intend to first complete the 0 min column, then continue with the best result based on the F1-score, and complete this row for different prediction lead times. Thereafter, we will complete the full column for the best prediction lead time. Continuing interchanging between completing columns and rows, until everything is complete, or we run out of time.}
\label{tab:model_runs}
\end{table}

\section{Discussion}
Results from our deep learning model will be compared to results from the traditional machine learning model presented in \cite{Line_resultater}. Our model will be tested on the Wrist Angel data both as pre-trained with and without fine-tuning and as non-pre-trained, to examine if pre-training and fine-tuning improves performance. The model will be evaluated with and without removing the observations classified as physical activity to investigate if a simple physical activity classifier helps in stress prediction. 
The signals will be shifted with different gaps between the end of the signal and the OCD event tag, such that we can explore the performance of predicting OCD events at different times ahead.
The Wrist Angel data is being split in windows of both 1min and 5min, to examine if the length of the time series influences the performance.

By publishing a statistical analysis plan before handling the Wrist Angel data, we ensure that the model and data parameters are set, such that, no data interpretation bias influences our results and decisions. 

The Wrist Angel dataset introduces some limitations. First, the limited number of participants, and thus the generalizability of the model can not be properly evaluated.  
Second, given the in-the-wild nature of the data, a significant portion will most likely be influenced by noise, for instance, because the wristband is not properly worn. 









\newpage
\bibliographystyle{unsrtnat}
\bibliography{bibliography} 

\newpage
\section{Appendix}

\subsection{Model Card \cite{model_cards}}
\label{app:model}


\newenvironment{mcsection}[1]
    {%
        \textbf{#1}

        \begin{itemize}[leftmargin=*,topsep=0pt,itemsep=-1ex,partopsep=1ex,parsep=1ex,after=\vspace{\medskipamount}]
    }
    {%
        \end{itemize}
    }

\begin{adjustwidth}{-30pt}{-30pt}
\begin{singlespace}

\tcbset{colback=white!10!white}
\begin{tcolorbox}[title=\textbf{Model Card - CNN Stress Classification},
    breakable, sharp corners, boxrule=0.7pt]

\small{

\begin{mcsection}{Model Details}
    \item Developed by Harald, Mia, and Simon at DTU, 2023 - Inspired by \cite{multiwave}.
    \item Convolutional Neural Network.
    \item Pre-trained for stress classification. Will be fine-tuned for obsessive-compulsive disorder OCD event classification in children and adolescents.
\end{mcsection}

\begin{mcsection}{Intended Use}
    \item Intended use is to predict OCD events.
    \item Intended users are children with OCD, wearing an Empatica E4 wristband, such that the model can predict and warn, before an OCD event.
    \item Not specifically pre-trained for either children or OCD events, hence could be used for several age groups and various acute stress episodes.
\end{mcsection}

\begin{mcsection}{Factors}
    \item Relevant factors that might affect the model performance are personal factors such as a person's age, health, shape, and environmental factors such as temperature, and humidity.
    \item Evaluation factors are 1: with or without physical activity classification, 2: using respectively 1min and 5min intervals, 3: varying the gap between the signal and the event tag (0 to 5min), and 4: the five different ways of applying the model as described in \cref{sec:DLmodel}. 
\end{mcsection}

\begin{mcsection}{Metrics}
    \item Accuracy and F1-score of stress classification are calculated.
    \item Metrics are reported as an average with a standard deviation of 10 runs.
\end{mcsection}

\begin{mcsection}{Training Data}
    \item The DTU dataset \\
    Physiological data from Empatica E4 from 28 subjects. The stress categorization comes from a timed puzzle task and the non-stress categorization from a resting phase.
    
    \item WESAD \cite{wesad} \\
    Physiological data from Empatica E4 from 15 subjects. The stress categorization comes from a Trier Social Stress Test (TSST) \cite{tsst} and the non-stress categorization from a resting phase and watching funny videos.
    
    \item AffectiveROAD dataset \cite{Road_dataset} \\
    Physiological data from Empatica E4 from 10 subjects (14 drives). The categorization comes from a subjective continuous stress evaluation of the driver during various driving scenarios.
    
    \item ADARP dataset \cite{adarp_dataset}\\
    Physiological data from Empatica E4 from 11 subjects diagnosed with alcohol use disorder (AUD). Data recorded in the wild. The stress categorization is from the subject tagging when they felt stressed. 
\end{mcsection}

\begin{mcsection}{Evaluation Data}
    \item Running evaluations of the model have been done by leaving one training data set out as test data. 
    \item The final evaluation will be done on a test set of the WristAngel data. 
    \item The WristAngel dataset \cite{wristangel_exp_protocol} \\
    Physiological data from Empatica E4 from 8 children with OCD. Data recorded in the wild. The stress categorization is from the child tagging when they felt stressed due to OCD symptoms.    
\end{mcsection}

\begin{mcsection}{Ethical Considerations}
    \item Physiological data as used for the model, is very personal and should be handled in accordance with GDPR. 
\end{mcsection}

\begin{mcsection}{Caveats and Recommendations}
    \item The final evaluation of the model has not been done yet, hence usefulness and considerations are subject to change.
\end{mcsection}

\textbf{Quantitative Analyses} 

To be written after testing on the WristAngel data.

} 
\end{tcolorbox}
\end{singlespace}
\end{adjustwidth}

\newpage

\subsection{Initial daily plan for WristAngel data analysis \\ (Week 50, 2023: Monday to Friday)}
\label{app:daily_plan}

Day 1 (Dec 11.): Understand the setup and how the data is formatted

Day 2 (Dec 12.): Data preprocessing (this includes generating the datasets according to our models)

Day 3 (Dec 13.): The initial in-clinic data for movement

Day 4 (Dec 14.) – Day 5 (Dec 15.): Training and evaluating using WristAngel data

Additionally, a 3-day buffer where we could be there sporadically.
\newpage

\subsection{Flow chart}
\begin{figure}[H]
\centering
\includegraphics[angle=90,origin=c,scale=0.52]{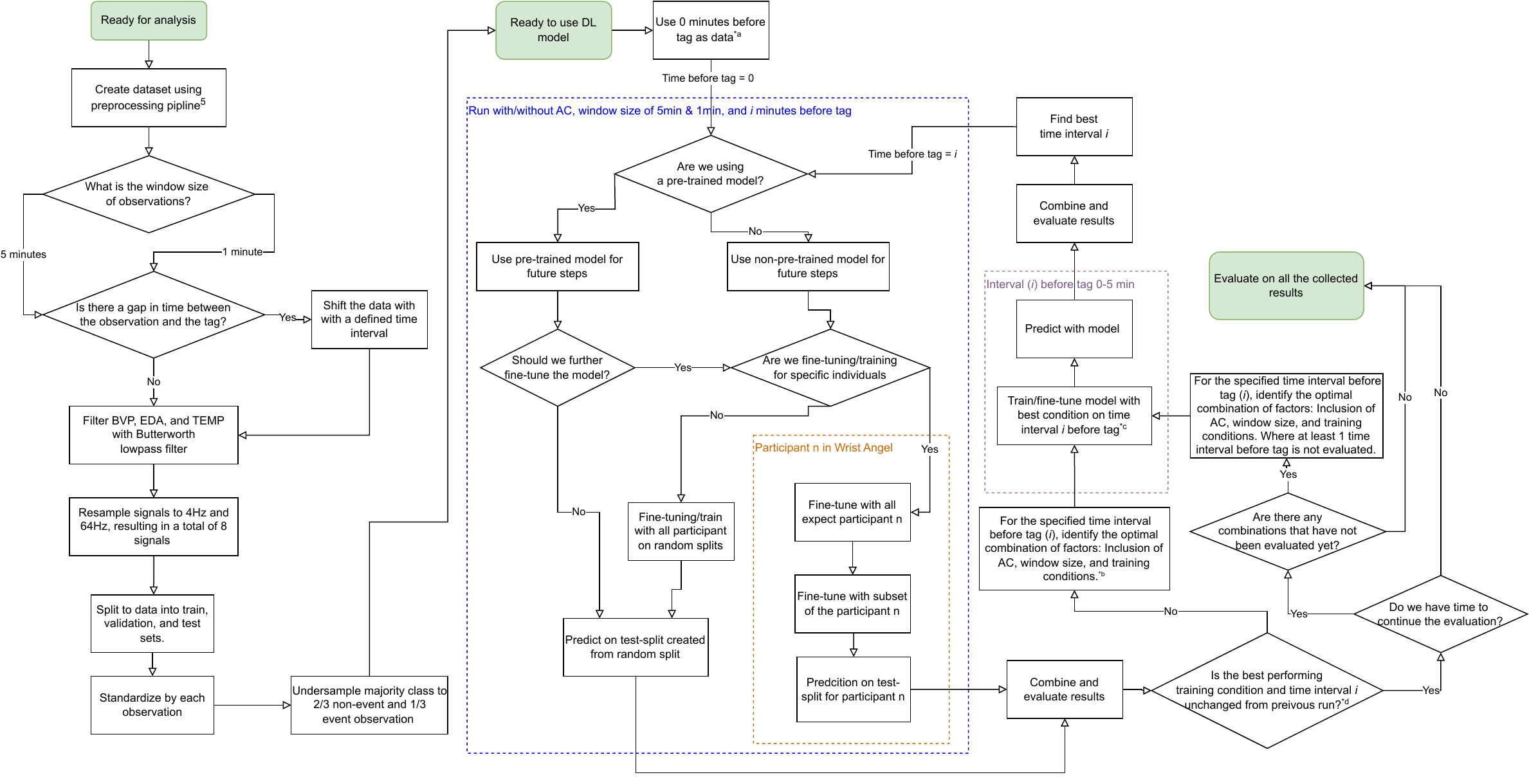}
    \label{fig:flowChart_appendix}
\end{figure}

\subsection{Example of how to evaluate model performance}
\begin{table}[h]
    \begin{subtable}[t]{0.45\textwidth}
\begin{tabular}{lllll}
\multicolumn{1}{c}{} & \multicolumn{4}{c}{Prediction time} \\
\multicolumn{1}{c}{\multirow{-2}{*}{\textbf{Condition 1}}} & 0 & 1 & $\cdots$ & 5 \\
Model 1 & \cellcolor[HTML]{FFFC9E} &  &  &  \\
Model 2 & \cellcolor[HTML]{FFFC9E} &  &  &  \\
\vdots & \cellcolor[HTML]{FFFC9E} &  &  &  \\ \hline
\multicolumn{1}{c}{} & \multicolumn{4}{c}{Prediction time} \\
\multicolumn{1}{c}{\multirow{-2}{*}{\textbf{Condition 2}}} & 0 & 1 & $\cdots$ & 5 \\
Model 1 & \cellcolor[HTML]{FFFC9E} &  &  &  \\
Model 2 & \cellcolor[HTML]{FFFC9E} &  &  &  \\
\vdots & \cellcolor[HTML]{FFFC9E} &  &  & 
\end{tabular}
\caption{Initial step: Each model in each condition with prediction lead time 0.}
\label{tab:TableFlow1}
\end{subtable}
\hfill
\begin{subtable}[t]{0.45\textwidth}
\begin{tabular}{lllll}
\multicolumn{1}{c}{} & \multicolumn{4}{c}{Prediction time} \\
\multicolumn{1}{c}{\multirow{-2}{*}{\textbf{Condition 1}}} & 0 & 1 & $\cdots$ & 5 \\
Model 1 & \cellcolor[HTML]{EFEFEF}0.2 &  &  &  \\
Model 2 & \cellcolor[HTML]{9AFF99}0.7 & \cellcolor[HTML]{FFFC9E}{\color[HTML]{FFFC9E} } & \cellcolor[HTML]{FFFC9E}{\color[HTML]{FFFC9E} } & \cellcolor[HTML]{FFFC9E}{\color[HTML]{FFFC9E} } \\
$\vdots$ & \cellcolor[HTML]{EFEFEF}$\vdots$ &  &  &  \\ \hline
\multicolumn{1}{c}{} & \multicolumn{4}{c}{Prediction time} \\
\multicolumn{1}{c}{\multirow{-2}{*}{\textbf{Condition 2}}} & 0 & 1 & $\cdots$ & 5\\
Model 1 & \cellcolor[HTML]{EFEFEF}0.3 &  &  &  \\
Model 2 & \cellcolor[HTML]{EFEFEF}0.3 &  &  &  \\
$\vdots$ & \cellcolor[HTML]{EFEFEF}$\vdots$ &  &  & 
\end{tabular}
\caption{Second step: Pick the row/model and condition with the best performance. Run this model for each prediction lead time.}
\label{tab:TableFlow2}
\end{subtable}\\
\begin{subtable}[t]{0.45\textwidth}
\begin{tabular}{lllll}
\multicolumn{1}{c}{} & \multicolumn{4}{c}{Prediction time} \\
\multicolumn{1}{c}{\multirow{-2}{*}{\textbf{Condition 1}}} & 0 & 1 & $\cdots$ & 5 \\
Model 1 & \cellcolor[HTML]{EFEFEF}0.2 &  &  &  \cellcolor[HTML]{FFFC9E}{\color[HTML]{FFFC9E} }\\
Model 2 & \cellcolor[HTML]{EFEFEF}0.7 & \cellcolor[HTML]{EFEFEF}0.4 & 
\cellcolor[HTML]{EFEFEF}$\cdots$ & \cellcolor[HTML]{9AFF99}0.75  \\
\vdots & \cellcolor[HTML]{EFEFEF}$\vdots$ &  &  & \cellcolor[HTML]{FFFC9E}{\color[HTML]{FFFC9E} } \\ \hline
\multicolumn{1}{c}{} & \multicolumn{4}{c}{Prediction time} \\
\multicolumn{1}{c}{\multirow{-2}{*}{\textbf{Condition 2}}} & 0 & 1 & $\cdots$ & 5 \\
Model 1 & \cellcolor[HTML]{EFEFEF}0.3 &  &  &  \cellcolor[HTML]{FFFC9E}{\color[HTML]{FFFC9E} }\\
Model 2 & \cellcolor[HTML]{EFEFEF}0.3 &  & &\cellcolor[HTML]{FFFC9E}{\color[HTML]{FFFC9E} }   \\
$\vdots$ & \cellcolor[HTML]{EFEFEF}$\vdots$ &  &  & \cellcolor[HTML]{FFFC9E}{\color[HTML]{FFFC9E} }
\end{tabular}
\caption{Thirds step: Pick column/prediction lead time with best performance. Run each model and condition with this prediction lead time.}
\label{tab:TableFlow3}
\end{subtable}
\hfill
\begin{subtable}[t]{0.45\textwidth}
\begin{tabular}{lllll}
\multicolumn{1}{c}{} & \multicolumn{4}{c}{Prediction time} \\
\multicolumn{1}{c}{\multirow{-2}{*}{\textbf{Condition 1}}} & 0 & 1 & $\cdots$ & 5 \\
Model 1 & \cellcolor[HTML]{EFEFEF}0.2 &  &  & \cellcolor[HTML]{EFEFEF}0.4 \\
Model 2 & \cellcolor[HTML]{EFEFEF}0.7 & \cellcolor[HTML]{EFEFEF}0.4  & \cellcolor[HTML]{EFEFEF}$\cdots$ & \cellcolor[HTML]{9AFF99}0.75\\
$\vdots$ & \cellcolor[HTML]{EFEFEF}$\vdots$ &  &  &\cellcolor[HTML]{EFEFEF}$\vdots$  \\ \hline
\multicolumn{1}{c}{} & \multicolumn{4}{c}{Prediction time} \\
\multicolumn{1}{c}{\multirow{-2}{*}{\textbf{Condition 2}}} & 0 & 1 & $\cdots$ & 5 \\
Model 1 & \cellcolor[HTML]{EFEFEF}0.3 &  &  &  \cellcolor[HTML]{EFEFEF}0.3\\
Model 2 & \cellcolor[HTML]{EFEFEF}0.3 &  &  &  \cellcolor[HTML]{EFEFEF}0.6\\
$\vdots$ & \cellcolor[HTML]{EFEFEF}$\vdots$ &  &  & \cellcolor[HTML]{EFEFEF}$\vdots$
\end{tabular}
\caption{Final step: Performance has not improved.}
\label{tab:TableFlow4}
\end{subtable}
\hfill
\caption{The steps of how we will evaluate the models are shown. The yellow cells are steps that will be taken. The green cells show the combination that created the best results and is the one we will continue. After the final step if time allows, we will continue with the best non-filled row or column.}
 \label{tab:FlowTable}
\end{table}







\end{document}